\documentclass[preprint,12pt]{elsarticle}
\usepackage[margin=3cm]{geometry}

\usepackage{amsmath,amssymb,amsfonts}
\usepackage{algorithmic}
\usepackage{graphicx}
\usepackage{subfigure}
\usepackage[version=4]{mhchem}
\usepackage{xcolor}
\usepackage{textcomp}
\usepackage{hyperref}
\usepackage{newunicodechar}
\usepackage{setspace}
\usepackage{nomencl}
\makenomenclature
\usepackage{etoolbox}
\usepackage[ruled,vlined]{algorithm2e}
\usepackage{tcolorbox}
\usepackage{multirow}
\tcbuselibrary{breakable}
\usepackage{float}
\usepackage{placeins}
\usepackage{booktabs}
\usepackage{xurl}
\newunicodechar{ϕ}{\phi}
\newunicodechar{₂}{$_2$}
\newunicodechar{ψ}{\psi}

\newtcolorbox{algobox}{
  colback=white, colframe=black,
  boxrule=0.8pt, left=4pt, right=4pt,
  top=4pt, bottom=4pt, arc=0pt,
  breakable
}

\journal{Electric Power Systems Research}

\begin{document}

\begin{frontmatter}

\title{Decentralized Operations of Decarbonized Chemical Plants with Renewable-driven Transmission Systems}

% \fnref{equal} marks the equal contribution footnote
% \corref{cor1} marks the corresponding author
\author[inst1]{Richard Reed\fnref{equal}}
\ead{ricky.reed@okstate.edu}

\author[inst1]{Kazi Arman Ahmed\fnref{equal}}
\ead{kazi.ahmed@okstate.edu}

\author[inst2]{Saba Ghasemi}
\ead{saba.ghasemi_naraghi@okstate.edu}

\author[inst2]{Zheyu Jiang}
\ead{zheyu.jiang@okstate.edu}

\author[inst1]{Paritosh Ramanan\corref{cor1}}
\ead{paritosh.ramanan@okstate.edu}

\cortext[cor1]{Corresponding author}
\fntext[equal]{These authors contributed equally to this work.}

\affiliation[inst1]{organization={School of Industrial Engineering and Management, Oklahoma State University},
            city={Stillwater},
            postcode={74078},
            state={OK},
            country={USA}}
\affiliation[inst2]{organization={School of Chemical Engineering, Oklahoma State University},
            city={Stillwater},
            postcode={74078},
            state={OK},
            country={USA}}

\begin{abstract}
Electrification of steam cracking offers a promising pathway to industrial decarbonization, provided that the electricity is sourced from renewable energy. However, integrating electrified chemical plant microgrids with a decarbonized power grid requires joint operations planning between Independent System Operators and chemical plants, which is hindered by the highly confidential nature of plant operational data. In this paper, we propose a privacy-friendly decentralized framework based on data isolation that jointly optimizes the Unit Commitment problem in the power system and microgrid scheduling in electrified ethylene plants. The framework employs the Alternating Direction Method of Multipliers (ADMM), augmented with an auxiliary system-level penalty that accelerates convergence, allowing each subsystem to solve its local subproblem and share only minimal coordination signals. To reflect real-world conditions, numerical experiments are conducted on the ACTIVSg2000 test case, a synthetic model of the Texas transmission network, with 26 steam cracking plants identified from Texas mapped to their nearest grid connection points. In doing so, we characterize the cost of privacy-friendly decomposition on joint power and chemical system decisions, showing that data isolation results in consistently small optimality gaps, and that its emissions consequences are load-dependent and non-monotone.
\end{abstract}

\begin{keyword}
Decentralized optimization \sep Unit commitment \sep ADMM \sep Privacy-friendly computation \sep Industrial electrification \sep Renewable energy integration
\end{keyword}

\end{frontmatter}

\section{Introduction}
The U.S. manufacturing sector accounts for approximately 20\% of the nation's total primary energy consumption and greenhouse gas (GHG) emissions \cite{doe}. With the shift in the U.S. energy landscape towards renewables, chemical process heating is also actively seeking electrification \cite{dharik}. Therefore, it is estimated that for large-scale electrification of chemical process heating, a robust renewable-driven power grid integration framework would be imperative \cite{AgrawalSiirola2023}. The viability of such integrations depends on a privacy-friendly data sharing framework between Independent System Operators (ISOs) \cite{rto} and chemical plants, which can be used to drive joint chemical and power system planning and operations in real-time. Unfortunately, the highly confidential nature of plant operation data can serve as the pivotal roadblock for data sharing between chemical and power system stakeholders which can derail successful grid integration efforts geared towards large-scale decarbonization \cite{privacy}. Therefore, in this paper, we develop a privacy-friendly joint chemical and power system operations planning paradigm that eliminates the need to move data from the plants.

Successfully integrating chemical and power system stakeholder operations must consider two distinct operational aspects. First, process heating units in chemical plants are required to operate at a steady state \cite{Naraghi2025}. As a result, their energy demand needs must be met by a rapidly evolving power system comprising a growing share of variable renewable energy, which is intermittent and volatile in nature \cite{eia}. Second, due to the high energy demand of electrified chemical process heating, power system stakeholders, such as Independent System Operators (ISO) and chemical plants, must jointly optimize their operations on a near real-time basis. In order to do so, chemical plants must relinquish ownership of their localized operational constraints and datasets to enable joint planning \cite{Naraghi2025,escape_25_microgrid}. Therefore, a robust, real-time framework to facilitate \textit{information sharing} between power system stakeholders and chemical plants is an essential precursor for joint integrated planning framework.

However, enabling a robust information sharing paradigm is hampered by the lack of privacy-friendly alternatives for joint operations that are capable of high solution quality while enabling stakeholders to retain full ownership of their local data. Additionally, from a planning perspective, there is limited research in modeling electrified process heating demand interdependencies across multiple plants in tandem with renewable electricity generation from power systems. These gaps are particularly evident in regions like Texas, which hosts a disproportionate share of the nation's chemical manufacturing capacity \cite{crackingsurvey} along with one of the largest and most renewable-rich grids in the country. As a result, in this paper, we develop decentralized mechanisms for integrating transmission system production scheduling with chemical plant operations so as to alleviate privacy concerns among stakeholders.

Conventionally, the optimal production scheduling for large scale transmission network is carried out using the Unit Commitment (UC) optimization problem to meet varying electricity demands \cite{padhy2004survey_uc}. The UC problem solution provides optimal decisions regarding generator commitment and dispatch subject to network flow and security constraints while minimizing operational costs, improving system reliability, and satisfying power demand \cite{hpcuc,ostrowski2011tight}. Historically, UC formulations consider conventional grids dominated by thermal generators \cite{ostrowski2011tight}. However, new challenges for the UC problem arise as the energy landscape evolves with a growing emphasis on industrial decarbonization coupled with the integration of renewable energy resources \cite{dharik,hpcuc}.

On the other hand, chemical manufacturing remain significant contributors to global carbon emissions. One of the classic examples is ethylene production via steam cracking. Ethylene crackers, which require substantial energy to heat chemicals for production, account for approximately 2.5\% of total energy consumption in the U.S., leading to high levels of \ce{CO2} emissions \cite{roadmap}. Electrifying these processes represents a promising pathway to decarbonization, provided that the electricity used is sourced from renewable energy \cite{Tijani2022,dharik,Balakotaiah2022,RodriguezGil2025}.

However, realizing an electrified future for ethylene cracking requires overcoming the privacy and information sharing bottleneck between centralized grid planning and distributed industrial cracking operations \cite{cisareport}. As a result, in this paper, we develop a decentralized optimization formulation for joint cracking and transmission system planning to resolve privacy bottlenecks. Our decentralization strategy allows chemical plants to interact with power grids without disclosing proprietary or sensitive information, while still enabling effective integration with renewable energy resources. By leveraging decentralized, consensus-based optimization paradigms \cite{kargarian2015distributed_scuc,ramanan2017asynchronous,ramanan2019asynchronous,ramanan2021large},  such as the Alternating Direction Method of Multipliers (ADMM) \cite{javad,xavier2024decomposable}, stakeholders can achieve a balance between operational privacy and system-wide optimization. 

Consequently, our approach relies on the ADMM framework to decouple transmission system and chemical plant operations at the bus level. Instead of centralized aggregation of cracker constraints, our decentralized approach enables joint planning solely through the iterative ADMM driven balancing of chemical process heating demand estimates between the transmission system and chemical plants. In order to demonstrate the efficacy of our proposed approach, we curate a large scale transmission system and chemical plant experimental case study. We demonstrate using rigorous computational results, that our decentralized methodology delivers similar solution qualities as centralized planning approaches without the need to move localized cracker constraints and data. Additionally, using a post-plan analysis  we show that a decentralized mechanism delivers very similar emissions estimates relative to the centralized method under varying transmission demand. We summarize our key contributions as follows:
\begin{itemize}
    \item A novel formulation of the UC problem that incorporates the unique energy demands of chemical plants alongside renewable energy resources.
    \item A decentralized joint optimization approach using ADMM to ensure privacy while achieving efficient coordination between chemical plants and power grids.
    \item A geographically grounded case study coupling 26 ethylene plants identified from Texas with the ACTIVSg2000 synthetic Texas transmission network, through which we show that data isolation results in consistently small optimality gaps while its emissions consequences are load-dependent and non-monotone.
    \item An empirical study of CO\textsubscript{2} emission implications as a consequence of decentralized formulation choices for varying demand cases.
\end{itemize}
Through these contributions, this work demonstrates a concrete, regionally grounded pathway for integrating ethylene plants into a renewable-driven power grid, offering a replicable framework for industrial and energy sector decarbonization more broadly. The remaining sections in this paper are structured as follows. Section~\ref{sec:related} discusses related works, focusing on prior research on decentralized optimization in the unit commitment (UC) problem, particularly those involving Lagrangian techniques and the ADMM framework. Section~\ref{sec:decentralized_operations} presents the decentralized operations formulation, detailing the integration of chemical plants into power grids and describing the iterative optimization process using ADMM. Section~\ref{sec:methodology} elaborates on the decentralized solution methodology, with emphasis on the ADMM-based approach, its implementation, and the communication protocol between plants and the power system. Section~\ref{sec:results} provides experimental results, comparing the decentralized model against a centralized variant for benchmarking. It also includes a comprehensive analysis of the impact of decarbonized grids on \ce{CO2} emissions, accounting for the integration of renewable energy resources. Finally, Section~\ref{sec:conclusion} concludes the paper by summarizing the key findings, highlighting the advantages of decentralized approaches, and suggesting directions for future research.

\section{Related Works} \label{sec:related}
The Unit Commitment (UC) problem has long been a cornerstone of power system research, focusing on optimizing generator schedules to meet demand while minimizing costs and ensuring reliability. Foundational methodologies were established by early surveys such as Padhy \cite{padhy2004survey_uc}, which outlined the evolution of UC techniques. As grids have shifted toward renewable energy integration, the complexity of UC models has increased significantly, particularly in addressing the stochastic nature of variable generation. To manage this volatility, researchers like Wang et al. \cite{wang2008scuc_wind} and Yang et al. \cite{yang2022review_scu_commitment} developed security-constrained UC models that explicitly account for wind and solar variability, providing robust frameworks for modern, decarbonizing grids.

To address the computational challenges posed by large-scale UC problems, decentralized and distributed optimization techniques have emerged as powerful alternatives to centralized solvers. Kargarian et al. \cite{kargarian2015distributed_scuc} introduced a distributed framework for security-constrained UC, demonstrating the scalability of decentralized methods in managing complex power systems. Subsequent works by Ramanan et al. \cite{ramanan2021large,ramanan2019asynchronous, ramanan2017asynchronous} further advanced this field by proposing asynchronous and differentially private decentralized solution frameworks, highlighting their effectiveness in handling large-scale systems with varying degrees of information sharing and privacy constraints.

A primary driver of these decentralized approaches is the Alternating Direction Method of Multipliers (ADMM), which offers strong convergence guarantees and flexibility in problem decomposition. The theoretical groundwork for ADMM was laid by Boyd et al. \cite{boyd2011admm_foundations} for unconstrained problems and extended by Wei and Ozdaglar \cite{wei2012distributed_admm_cdc} to constrained optimization. Recent advancements by Biswas et al. \cite{biswas2023decentralized_opf} and Erseghe \cite{erseghe2014distributed_opf_admm} demonstrated the potential of ADMM for handling decentralized operations in systems with high renewable penetration. Furthermore, specific applications of ADMM-based decompositions to the UC domain have been well-explored \cite{ xavier2024decomposable,ramanan2019asynchronous,javad}, establishing it as a reliable paradigm for scalable grid optimization.

Despite these advances, existing literature largely focuses on the technical aspects of decentralized UC as a means to drive operational and privacy benefits of decentralization. However, none of these works explicitly characterize the cost of decomposition choices on solution quality and emissions outcomes in coupled power-chemical systems. This paper addresses that gap by building upon these foundational works to quantify the privacy-optimality tradeoff. Specifically, we provide the first systematic empirical characterization of how data-isolating ADMM decomposition affects both economic performance and environmental impact when chemical plant microgrids are integrated with decarbonized power grids.

\section{Decentralized Joint Operations Planning} \label{sec:decentralized_operations}

\begin{figure}[ht!]
    \centering
    \includegraphics[width=0.95\textwidth, trim=25pt 50pt 25pt 50pt, clip]{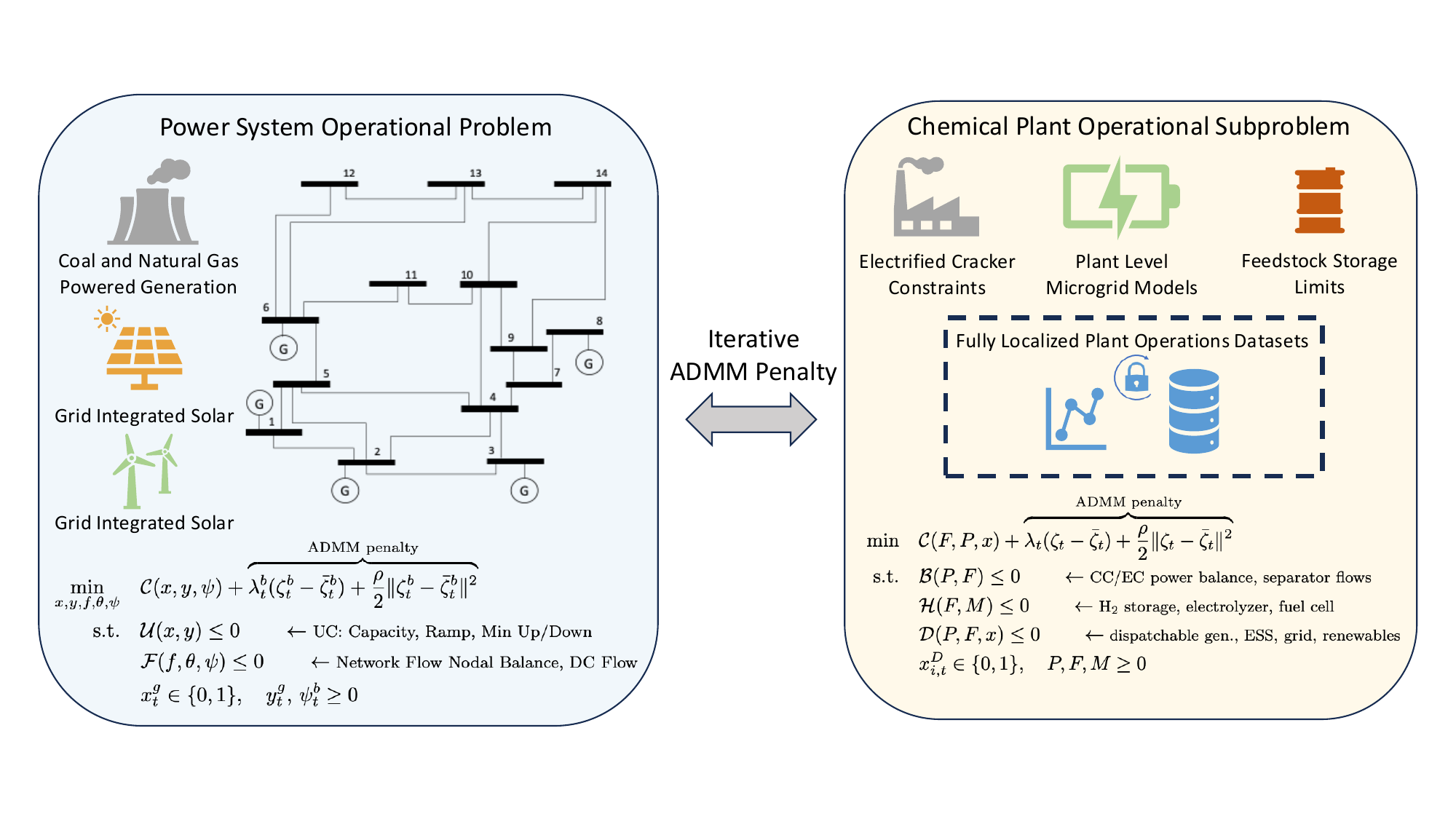}
    \caption{Decentralized microgrid superstructure for steam cracking. The power system acts as the central coordinator exchanging consensus variables with chemical plants connected at specific buses.}
    \label{fig:ps-cp-overarching}
\end{figure}

In this section, we present the decentralized problem formulation for two distinct systems, derived from the ADMM framework \cite{boyd2011admm_foundations, ramanan2019asynchronous, biswas2023decentralized_opf, javad}. This approach aims to integrate chemical plants into the power system by distributing each plant and connecting to the network at the bus level. The power system acts as the main process and exchanges a consensus variable with each plant iteratively. A new iteration occurs once the main process receives each consensus variable from the plants and continues until global optimization is reached. This decentralized formulation provides a framework that enables the integration of two separate processes and offers applicability in real-world situations that would otherwise be infeasible. The overarching view of our proposed decentralized framework that jointly integrates electrified cracker demand with the power system is represented in Figure \ref{fig:ps-cp-overarching}.
We present the decentralized UC formulation in Problem \eqref{eq:PS_UC}, with constraints described below.

\subsection{Power System Formulation}
\noindent The fundamental UC problem formulation is represented by Problem \eqref{eq:PS_UC}.
\begin{equation}\label{eq:PS_UC}
\begin{aligned}
    \mathcal{L}_{r} = & \sum_{t\in T} \Big[\sum_{g\in G} d^g y_t^g + c^g x_t^g  + \nu\sum_{b\in B} \psi^b_t \Big]
    %\\
    %& + \sum_{u \in \mathcal{B}} \sum_{t \in T} \left[ \lambda^{g}_{t,u} (\zeta^u_t - \overline{\zeta}^u_t) + \frac{\rho}{2} (\zeta^u_t - \overline{\zeta}^u_t)^2 \right]
    %\\
    % & +  \sum_{t \in T} \left[\lambda_t(\psi_t) + \frac{\rho}{2} ||\psi_t||^2\right]
\end{aligned}
\end{equation}
Problem \eqref{eq:PS_UC} specifically pertains to a transmission system where the set of all buses is denoted by $B$, and $T$ represents the operational planning horizon. Each bus $b \in B$ may have associated generators and chemical plants, belonging to the sets $G$ and $P$, respectively. The set $B^u \subseteq B$ contains all buses connected to a given bus $b$. For each generator $g \in G$, $y_t^g$ denotes its electricity dispatch, $x_t^g \in \{0,1\}$ indicates its commitment state, and $\eta_t^g$ denotes renewable power generation. For each bus $b$, $\nu,\psi^b_t$ represents the localized demand curtailment and its cost coefficient respectively. The variables $\pi^g_{Ut}$ and $\pi^g_{Dt}$ represent up and down reserve deployment, while $\theta_t^b$ is the phase angle at bus $b$, and $\lambda_t^b$ is the Lagrangian multiplier enforcing nodal power balance, considering power system demand and chemical process heating requirement. Power flow from bus $u$ to $b$ is represented by $f_t^{bu}$. The parameters $c^g$ and $d^g$ denote commitment and dispatch costs for generator $g$, while $P^g_{\min}$ and $P^g_{\max}$ specify its generation limits. The minimum up and down times are $M_U^g$ and $M_D^g$, and the ramping limit is $R^g$. The demand at bus $b$ is given by $\delta_t^b$ while transmission line capacity is given by $F_{\max}^{bu}$ with $\Gamma^{bu}$ denoting the phase-angle-to-power-flow conversion constant for $bu$. For integrated chemical processing and electricity operations, we let, $\zeta_t^b$ denotes cracker-related electrical demand at bus $b$, while $\tilde{\zeta}_{t}^p$ is the electricity demand of chemical plant $p \in P$. 

%The term $\dot{\zeta}_t^p$ represents a penalty associated with plant $p$, and $\bar{\zeta}_t^b$ denotes the average cracker demand at bus $b$. The term $\lambda_t^b$ also appears as the Lagrangian penalty associated with cracker-demand constraints. 

Finally, we incorporate operational constraints \eqref{eq:2a}-\eqref{eq:2g} for production scheduling at the power system stakeholder. Constraint \eqref{eq:2a} ensures that production at each generator is bounded by its minimum and maximum capacity.
\begin{equation}
    \begin{aligned}
        \quad P^g_{min}x^g_t \leq y^g_t \leq P^g_{max}x^g_t\quad \forall t\in T,\, \forall g \in G
    \end{aligned}
    \label{eq:2a}
\end{equation}
Constraints \eqref{eq:2b1}-\eqref{eq:2b3} enforce the minimum up and down time for each generator.
\begin{subequations}\label{eq:2b}
    \begin{align}
        \quad -\pi^g_{Dt} \leq x^g_t - x^g_{t-1} \leq \pi^g_{Ut}, \quad \forall t\in [2,T], \forall g \in G \label{eq:2b1}\\
        \quad \sum_{\forall i \in U_t} \pi^g_{Ui} \leq x^g_t \leq 1 - \sum_{\forall i \in D_t} \pi^g_{D_t}, \quad \forall t \in T,\, \forall g \in G\label{eq:2b2}\\
        U_t = [t - M^g_U + 1,t], D_t = [t - M^g_D + 1,t]\label{eq:2b3}
    \end{align}
\end{subequations}
Constraint \eqref{eq:2c} ensures generators adhere to respective ramping limitations.
\begin{equation}
    \begin{aligned}
        \quad -R^g \leq y^g_t - y^g_{t-1} \leq R^g, \quad \forall t \in [2, T], \forall g \in G
    \end{aligned}
    \label{eq:2c}
\end{equation}
Equation \eqref{eq:2d} ensures that flow across links is a function of the respective phase angles.
\begin{equation}
    \begin{aligned}
        \quad \Gamma^{bu}(\theta^b_t - \theta^u_t) = f^{bu}_t \quad \forall b \in B, \forall u \in B^u, \forall t \in T
    \end{aligned}
    \label{eq:2d}
\end{equation}
Constraint \eqref{eq:2e} ensures that at each bus $b$ the demand is satisfied by either the demand curtailment $\psi^b_t$, the connected thermal or renewable generator set $G^b$ or through net power inflow into bus $b$.
\begin{equation}
\begin{aligned}
    \sum_{ g \in G^b} (y^{g}_t + \eta^{g}_t) - (\delta^b_t + \zeta^b_t) +\psi^b_t
    &= \sum_{ \forall u \in B^b} \left[ \Gamma^{bu} (\theta^b_t - \theta^u_t) \right], \forall t \in T
\end{aligned} \label{eq:2e}
\end{equation}
% Equations \eqref{eq:2f}, \eqref{eq:2g} enforce global network flow constraints and compute demand curtailment respectively.

Equations \eqref{eq:2f}, enforces global network flow constraints.
\begin{equation}
    \begin{aligned}
        -F^{bu}_{max} \leq \Gamma^{bu}(\theta^b_t - \theta^u_t) 
        &\leq F^{bu}_{max}, \forall b \in B ,  \forall u \in B^u, \forall t \in T
    \end{aligned} \label{eq:2f}
\end{equation}
% \begin{equation}
%     \begin{aligned}
%         \psi_t = \left[ \sum_{b \in B} \delta_t^b + \sum_{p \in P} \zeta_t^p \right] - \left[ \sum_{g \in G} (y^g_t +  \eta^{g}_t) \right] \quad \forall t \in T
%     \end{aligned}\label{eq:2g}
% \end{equation}

% \begin{equation}
%     \begin{aligned}
%         \psi_t = \left[ \sum_{g \in G} (y^g_t +  \eta^{g}_t) \right] - \left[ \sum_{b \in B} \delta_t^b + \sum_{p \in P} \zeta_t^p \right] \quad \forall t \in T
%     \end{aligned}\label{eq:2g}
% \end{equation}

We now discuss the chemical plant operations optimization formulation with active power system integrations.

\subsection{Electrified Chemical Plant Formulation}
\begin{figure}[ht!]
    \centering    
    \includegraphics[width=0.9\textwidth]{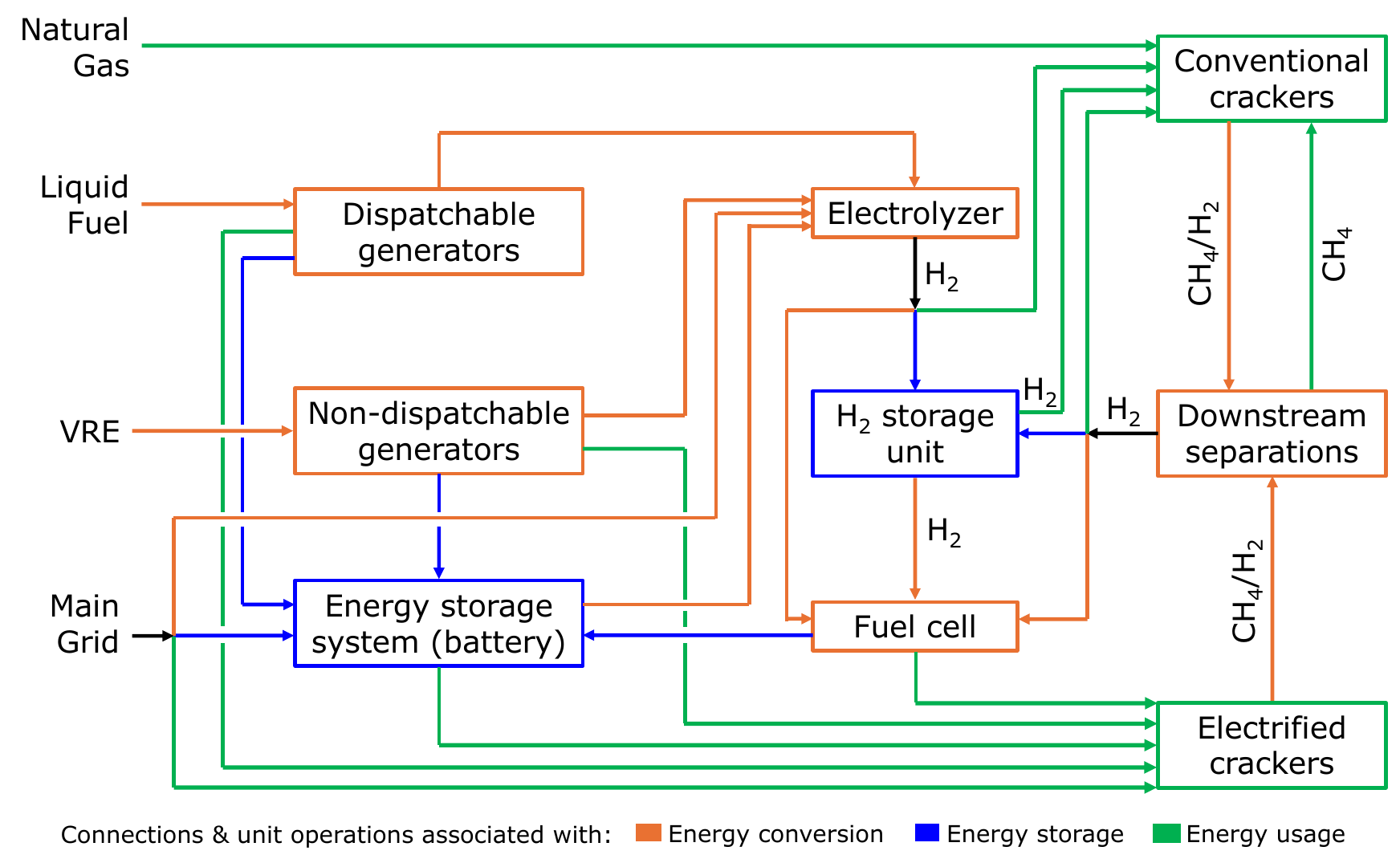}
    \vspace{-1em}
    \caption{Our envisioned microgrid superstructure for using electricity to supply process heat for steam cracking, which is adopted from our earlier work \cite{Naraghi2025}. Diverse energy sources supply heat for both electrified and conventional crackers that are present in the superstructure. Depending on the nature of the energy carriers, the connections shown in the superstructure can represent either energy or mass flows.}\label{fig:microgrid}
\end{figure}

The electrified chemical plant formulation consists of a mixed-integer formulation that determines the optimal operation of a microgrid associated with a chemical plant. In particular, we focus on an important application of electrification of steam cracking process for sustainable ethylene production \cite{Naraghi2025}. Ethylene is a platform chemical that serves as the key precursor for plastics, textiles, antifreeze, detergents, adhesives, rubber, and food packaging materials \cite{afpm}. The global annual production rate of ethylene exceeded 200 million metric ton (MMT) in 2022 and is expected to grow by more than 60\% by 2034 \cite{Precedence2024}. Ethylene is predominantly produced through steam cracking, an energy-intensive process that cracks hydrocarbon feedstocks such as ethane, propane, and naphtha at high temperatures in furnaces (known as cracker) \cite{Zimmermann2009}. The furnace heat is supplied by the combustion of fresh natural gas fuel or the methane fraction of the cracking byproduct, making steam cracking one of the most energy and emission-intensive chemical processes and accounting for 8\% of its total primary energy consumption \cite{REN2006425}. As a result, active research is underway to electrify steam cracking process. Our proposed model considers future ethylene plant as a microgrid (see Figure \ref{fig:microgrid}), whose process heating demand is provided by diverse renewable and non-renewable sources to support a continuous and steady state operation \cite{Naraghi2025,saba_centralized}. 

The model jointly optimizes the microgrid's scheduling decisions and the degree of electrification for the cracking process. It captures the interplay between electricity generation, storage, and process demand to enable cost-effective and emission-conscious operational planning. We present the microgrid scheduling formulation at plant $p$ as depicted in Problem \eqref{eq:microgrid_obj}, with objective $\mathcal{L}_p$ that needs to be minimized subject to constraints \eqref{eq:cc_power_balance}-\eqref{eq:ess_constraints}. 
\begin{equation}\label{eq:microgrid_obj}
    \begin{aligned}
    \mathcal{L}_p = & \sum_{t \in T} \Bigg[
    \text{c}_\text{NG}^\text{Fuel} \cdot F^\text{NG}_{\text{CC,}t} + \sum_{g \in G} \sum_{i \in I_g} \Big(c_g^\text{Fuel} F_{i, t}^D + c_g^\text{D} P_{i, t}^{D} + c^D_{\text{SU}, g} \max\{0, x_{i, t}^{D} - x_{i, t-1}^{D}\}  \\
    & \qquad \qquad + c^D_{\text{SD}, g} \max\{0, x_{i, t-1}^{D} - x_{i, t}^{D}\} \Big) + c_{t}^\text{G} P_{t}^\text{G} + c_t^\text{FC} P_{t}^\text{FC} + c_t^\text{EL} F_{EL, t}^{\text{H}_2} + c_t^\text{HS} M_{HS, t}^{\text{H}_2}
    %\\
    % & \qquad \qquad +
    % \left[ \lambda_t \cdot (\dot{\zeta}_t - \tilde{\zeta}_t) + \frac{\rho}{2} (\dot{\zeta}_t - \tilde{\zeta}_t)^2 \right] \\
    % &\qquad \qquad +
    % \left[ \lambda_t \cdot (q_t - \tilde{\zeta}_t) + \frac{\rho}{2} (q_t - \tilde{\zeta}_t)^2 \right]
\Bigg]
    \end{aligned}
\end{equation}
Constraint \eqref{eq:cc_power_balance} ensures that the conventional cracker (CC) units are powered through a mixture of natural gas, methane and hydrogen recovered from separation processes, and hydrogen supplied either from electrolyzers or from hydrogen storage units. Constraint \eqref{eq:ec_power_balance} specifies that electrified crackers (EC) satisfy their power demand using contributions from dispatchable units, non-dispatchable renewable resources, grid imports, energy storage, and fuel cell systems.
\begin{subequations}\label{eq:power_balance}
    \begin{align}
        %cc power balance
        P_\text{CC} =& \; \dot{Q}_\text{NG} F_{\text{CC,}t}^\text{NG} + \dot{Q}_{\text{CH}_4} F_{\text{sep, CC,}t}^{\text{CH}_4} + \dot{Q}_{\text{H}_2} \Big(F_{\text{sep, CC,}t}^{\text{H}_2} + F_{\text{EL,CC,}t}^{\text{H}_2} +  F_{\text{HS,CC,}t}^{\text{H}_2}\Big), \forall t \in T \label{eq:cc_power_balance}\\
        %ec power balance
        P_\text{EC} =& \; \sum_{g \in G} P_{\text{EC,}g,t}^\text{D} + P_{\text{EC,}t}^\text{ND} + P_{\text{EC,}t}^\text{G} + P_{\text{EC,}t}^\text{ESS} + P_{\text{EC,}t}^\text{FC}, \forall t \in T \label{eq:ec_power_balance}
    \end{align}
\end{subequations}
Constraint set \eqref{eq:sep_flow} governs post-cracking separation processes, ensuring proper allocation of methane and hydrogen byproducts across multiple pathways. 
\begin{subequations}\label{eq:sep_flow}
    \begin{align}
        %
        %separator flow 1
        F_{\text{sep,CC,}t}^{\text{CH}_4} =& r_{\text{CH}_4\text{,sep}} f_{\text{CH}_4}\Big(F_\text{CC}^{\text{CH}_4\text{/H}_2} + F_\text{EC}^{\text{CH}_4\text{/H}_2}\Big)\label{eq:sepflow_1}\\
        %
        %separator flow 2
        F_{\text{sep,CC,}t}^{\text{H}_2} +& F_{\text{sep,HS,}t}^{\text{H}_2} + F_{\text{sep,FC,}t}^{\text{H}_2}
                                         = r_{\text{H}_2\text{,sep}}\Big(1 - f_{\text{CH}_4}\Big) \Big(F_\text{CC}^{\text{CH}_4\text{/H}_2} + F_\text{EC}^{\text{CH}_4\text{/H}_2}\Big), \forall t \in T \label{eq:sepflow_2}
    \end{align}
\end{subequations}
The hydrogen mass balance and state evolution within the storage units are enforced through \eqref{eq:h2_storage}.
\begin{subequations}\label{eq:h2_storage}
\begin{align}
        &M_{\text{HS},t-1}^{\text{H}_2}\;+\;\Big(F_{\text{sep,HS},t}^{\text{H}_2}+ F_{\text{EL,HS},t}^{\text{H}_2} - F_{\text{HS,FC},t}^{\text{H}_2}-F_{\text{HS,CC},t}^{\text{H}_2}\Big)\Delta t\;=\;M_{\text{HS},t}^{\text{H}_2}, \label{eq:h2_storage_1}\\
        &F_{\text{EL,FC},t}^{\text{H}_2}\;+\;F_{\text{EL,HS},t}^{\text{H}_2}+ F_{\text{EL,CC},t}^{\text{H}_2}\;=\;F_{\text{EL},t}^{\text{H}_2} \forall t \in T\label{eq:h2_storage_2}\\
        &0 \;\le\; M_{\text{HS},t}^{\text{H}_2}\;\le\; \text{HSC},\quad M_{\text{HS},0}^{\text{H}_2} = M_{\text{HS,start}}^{\text{H}_2} \quad \forall t \in T \label{eq:h2_storage_3}
\end{align}
\end{subequations}

Hydrogen production via PEM electrolyzers, powered by electricity sourced from dispatchable, nondispatchable, grid, and energy storage resources, is modeled in \eqref{eq:electrolyzer_bounds}. 
\begin{subequations}\label{eq:electrolyzer_bounds}
    \begin{align}
        F_{\text{EL,}t}^{\text{H}_2}
        &= \frac{\eta_{\text{EL}}}{P_{\text{H}_2}}
        \Bigg(\sum_{g \in G} P_{\text{EL,}g,t}^{\text{D}}
        + P_{\text{EL,}t}^{\text{ND}}
        + P_{\text{EL,}t}^{\text{G}}
        + P_{\text{EL,}t}^{\text{ESS}} \Bigg) \label{eq:electrolyzer_bounds1}\\
        0 \leq & F_{\text{EL,}t}^{\text{H}_2}
        \leq \text{ELC}, \forall t \in T \label{eq:electrolyzer_bounds2}
    \end{align}
\end{subequations}
Constraint \eqref{eq:fuel_cell} determines the operation of fuel cells, establishing the relationship between hydrogen inflows and electrical energy output.
\begin{subequations}\label{eq:fuel_cell}
    \begin{align}
        P_{t}^{\text{FC}}
        &= \eta_{\text{FC}} \dot{Q}_{\text{H}_2}
        \Big(
        F_{\text{HS,FC,}t}^{\text{H}_2}
        + F_{\text{EL,FC,}t}^{\text{H}_2}
        + F_{\text{sep,FC,}t}^{\text{H}_2}\Big) \label{eq:fuel_cell1}\\
        P_{t}^{\text{FC}}& x_{t w}^{FC}
        \leq P_{t}^{\text{FC}}
        \leq P_{t}^{\text{FC}} x_{t w}^{FC} \label{eq:fuel_cell2}\\
        P_{t}^{\text{FC}}
        &= P_{\text{EC,}t}^{\text{FC}}
        + P_{\text{ESS,}t}^{\text{FC}}, \forall t \in T \label{eq:fuel_cell3}
    \end{align}
\end{subequations}
Power distribution between grid-supplied and locally generated electricity is governed jointly by \eqref{eq:gas_turbine} and \eqref{eq:renewables_balance}.
\begin{equation}\label{eq:gas_turbine}
    \begin{aligned}
        P_{t}^{\text{G}}
        = P_{\text{ESS,}t}^{\text{G}}&
        + P_{\text{EC,}t}^{\text{G}}
        + P_{\text{EL,}t}^{\text{G}},& 0 \leq P_{t}^{\text{G}}
        \leq \overline{P}^{\text{G}}, \forall t \in T .
    \end{aligned}
\end{equation}
\begin{equation}\label{eq:renewables_balance}
    \begin{aligned}
        P_{\text{ESS,}t}^{\text{ND}}
        + P_{\text{EC,}t}^{\text{ND}}
        + P_{\text{EL,}t}^{\text{ND}}
        = P_{t}^{\text{WT}} + P_{t}^{\text{PV}}, \forall t \in T .
    \end{aligned}
\end{equation}
Constraint set \eqref{eq:dispatchable_gen} models fuel-based dispatchable generation units, capturing power balance relations, generation limits, ramping restrictions, and minimum up/down time requirements, paralleling the classical unit commitment constraints in \eqref{eq:PS_UC}.
\begin{subequations}\label{eq:dispatchable_gen}
    \begin{align}
        P_{i,t}^{\text{D}}
        &= \dot{Q}_g F_{i,t}^{\text{D}}, \quad \forall i \in I_g \label{eq:dispatchable_gen1}\\
        P_{g,t}^{\text{D,min}}
        &\leq P_{i,t}^{\text{D}}
        \leq \overline{P}_g^{\text{D}}, \quad \forall i \in I_g \label{eq:dispatchable_gen2}\\
        P_{g,t}^{\text{D}}
        &= \sum_{i \in I_g} P_{i,t}^{\text{D}}, \label{eq:dispatchable_gen3}\\
        P_{g,t}^{\text{D}}
        &= P_{\text{ESS,}g,t}^{\text{D}}
        + P_{\text{EC,}g,t}^{\text{D}}
        + P_{\text{EL,}g,t}^{\text{D}} \label{eq:dispatchable_gen4}\\
        P_{i,t}^{\text{D}}& - P_{i,t-1}^{\text{D}}
        \leq \text{RU}_g^{\text{D}}, \quad \forall i \in I_g \label{eq:dispatchable_gen5}\\
        P_{i,t-1}^{\text{D}} &- P_{i,t}^{\text{D}}
        \leq \text{RD}_g^{\text{D}}, \quad \forall i \in I_g \label{eq:dispatchable_gen6}\\
        x_{i,t}^{\text{D}}& - x_{i,t-1}^{\text{D}}
        \geq 0, \\
        1 &- x_{i,t-1}^{\text{D}} + x_{i,t}^{\text{D}}
        \geq 0,  \forall g \in G, \; t \in T
    \end{align}
\end{subequations}
Finally, constraint set \eqref{eq:ess_constraints} defines the dynamics of the energy storage system (ESS), including power balance across charging and discharging modes and state-of-charge evolution.
\begin{subequations}\label{eq:ess_constraints}
    \begin{align}
        P_{C,t}^{\text{ESS}}
        &= P_{\text{ESS,}t}^{\text{ND}}
        + \sum_{g \in G} P_{\text{ESS,}g,t}^{\text{D}}
        + P_{\text{ESS,}t}^{\text{FC}}, \\
        E_{t}^{\text{ESS}}
        &= E_{t-1}^{\text{ESS}}
        - \Big(P_{C,t}^{\text{ESS}}
        - P_{DC,t}^{\text{ESS}}\Big) \Delta t, \\
        0 \leq E_{t}^{\text{ESS}}
        &\leq \text{ESC}, \quad
        E_{0}^{\text{ESS}} = E_{\text{start}}^{\text{ESS}}, \forall t \in T
    \end{align}
\end{subequations}

With the power system and chemical plant stakeholder models in place, we now present the ADMM based decentralized solution methodology for joint operations planning with data and constraint isolation. 

\section{Decentralized Solution Methodology} \label{sec:methodology}
Our joint decentralized operations optimization methodology is specifically geared towards addressing the optimization coupling challenges posed by the UC problem in \eqref{eq:PS_UC} and the chemical plant microgrid problem in \eqref{eq:microgrid_obj}. These represent two distinct but physically coupled systems (the transmission grid and the electrified chemical plants) that must coordinate under shared demand and resource constraints. As a result, our goal is to achieve high-quality solutions despite the coupling, while simultaneously preserving each plant’s operational privacy and minimizing the exchange of sensitive or proprietary information.

Our decentralized strategy is driven by the Alternating Direction Method of Multipliers (ADMM), in which each model is augmented with a penalty term and Lagrangian multipliers that enforce coordination through iterative updates. To further enhance robustness and improve solution quality, a second auxiliary system-level penalty is introduced which creates a demand target for plants based on system-wide unmet demand. The additional penalty structure helps guide convergence. %Computational advantages of this extended formulation are discussed in the following section. 

\subsection{Decentralized Power System Scheduling}
Our ADMM based methodology enables the decomposition of the global optimization problem into subproblems, which are solved iteratively and locally at each microgrid and power system component. This decentralized approach allows each plant to make operational decisions independently, with coordination achieved through iterative communication and dual variable updates. The essence of the ADMM framework lies in augmenting the objective function of each subproblem with a penalty term and a Lagrangian multiplier that penalizes discrepancies between shared variables, which, in this case, are the power demand at each bus. By iteratively updating local solutions and exchanging limited information, a global consensus can be achieved while maintaining data privacy. 

We begin with the conventional standard form ADMM-driven power system optimization objective as given by equation \eqref{eq:std-admm-ps}.
\begin{equation}\label{eq:std-admm-ps}
\min_{\zeta, f, \theta, x, y} \mathcal{L}_{r}(\overline{\zeta}, \lambda) = \mathcal{L}_{r} + \sum_{b \in B} \sum_{t \in T} \bigg[\lambda^{b}_{t} \left( \zeta_{t}^b - \overline{\zeta}_{t}^b \right) +\frac{\rho}{2} \left( \zeta_{t}^b - \overline{\zeta}_{t}^b \right)^2 \bigg]
\end{equation}
In equation \eqref{eq:std-admm-ps}, $\mathcal{L}_{r}$ represents the original power system objective function augmented by the ADMM penalty functions in which $\zeta_t^b$ represents the net cracker demand estimate at bus $b$ of the transmission system. The variable $\lambda^{b}_{t}$ is the Lagrangian multiplier associated with the nodal power balance constraint at bus $b$ during time step $t$. As a result, variable $\overline{\zeta}_t^b$ denotes the consensus estimate across the set of steam cracker units connected to bus $b$. The parameter $\rho > 0$ is the penalty coefficient that controls the weight of the consensus violation in the augmented Lagrangian.

In order to derive the consensus estimate $\overline{\zeta}_t^b$ at bus $b$, we begin by denoting the chemical plant microgrid demand estimate from plant $p$ as $\tilde{\zeta}_t^p$. Meanwhile the aggregated demand from all plants connected to bus $b$ denoted by $\hat{\zeta}_t^b$ can be computed through equation \eqref{eq:19a}.
\begin{equation}
\hat{\zeta}_t^b = \sum_{p \in P_b} \tilde{\zeta}_t^p, \quad \forall t \in T, \forall b \in B \label{eq:19a}
\end{equation}
Equation \eqref{eq:19a} ensures that the aggregated plant-level demands are accurately reflected at the bus level, forming a foundation for subsequent calculations. Next, equation \eqref{eq:19c} averages the plant demand and the power system demand values for a given bus.
\begin{equation}
\bar{\zeta}_t^b = \frac{\zeta_t^b + \hat{\zeta}_t^b}{2}, \quad \forall t \in T, \forall b \in B \label{eq:19c}
\end{equation}
The averaging step is critical in aiding the convergence of the ADMM framework, as it smoothens discrepancies between the plant and system-level demand values. Consequently, equation \eqref{eq:19d} updates the Lagrangian multiplier \( \lambda^{b}_{t} \) for each bus.
\begin{equation}
\lambda^{b}_{t} \gets \lambda^{b}_{t} + \rho \cdot \left( \zeta_t^b - \bar{\zeta}_t^b \right), \quad \forall t \in T, \forall b \in B \label{eq:19d}
\end{equation}
The lagrangian adjustment penalizes any deviation between the power system demand and the averaged demand, enforcing consistency across iterations and driving convergence to a feasible solution.

After each iterative solve of Problem \ref{eq:std-admm-ps}, the power system must communicate the bus-disaggregated estimate of cracker demand to each plant. In order to do so, we consider the ratio of the plant's demand with respect to the net cracker demand at the connected bus. Consequently, equation \eqref{eq:19b} can be used to compute each plant's cracker demand estimate from the viewpoint of the power system which can be duly communicated to the plant itself.
\begin{equation}
\dot{\zeta}_t^p = \frac{\Tilde{\zeta}_t^p}{\hat{\zeta}_t^b} \cdot \zeta_t^b, \quad \forall t \in T, \forall b \in B, \forall p \in P \label{eq:19b}
\end{equation}
By scaling this ratio using the updated power system cracker demand value, equation \eqref{eq:19b} ensures that the plant's contribution to the bus demand is proportionately adjusted. 

Equations \eqref{eq:19a}-\eqref{eq:19b} collectively govern the iterative updates in the ADMM framework, ensuring a balance between plant-level operations and bus-level demands while promoting efficient convergence through penalty and averaging mechanisms. However, to enhance the convergence speed and computational efficiency of the decentralized ADMM framework, a second auxiliary power system level penalty term is introduced. The power system penalty leverages the global power balance relationship to provide an estimate of system-wide unmet demand, which is used to influence the penalty applied to each chemical plant microgrid's load coordination. By doing so, the optimization formulation prioritizes demand fulfillment from the power system and enables more targeted penalization for imbalances, accelerating convergence.

The system-wide penalty term is driven by an estimation of unmet demand, defined at each time period \( t \in T \) as detailed in Equations \eqref{eq:2g} and \eqref{eq:20a}.
\begin{equation}
    \begin{aligned}
        \psi_t =  \left[ \sum_{b \in B} \delta_t^b + \sum_{p \in P} \zeta_t^p \right] - \left[ \sum_{b \in B} \sum_{g \in G^b} (y^g_t +  \eta^{g}_t) \right] = \sum_{b \in B} \psi_t^b \quad \forall t \in T
    \end{aligned}\label{eq:2g}
\end{equation}

\begin{equation}
    \bar{p}_t = \psi_t + \sum_{b \in B} \delta_t^b - \left[ \sum_{b \in B}\sum_{g \in G^b} (y^g_t +  \eta^{g}_t) \right]  \quad \forall t \in T.
    \label{eq:20a}
\end{equation}

% \begin{equation}
%     \bar{p}_t = \psi_t - \left[ \sum_{g \in G} (y^g_t +  \eta^{g}_t) \right] + \sum_{b \in B} \delta_t^b \quad \forall t \in T.
%     \label{eq:20a}
% \end{equation}

% \begin{equation}
%     \begin{aligned}
%         \psi_t = \left[ \sum_{g \in G} (y^g_t +  \eta^{g}_t) \right] - \left[ \sum_{b \in B} \delta_t^b + \sum_{p \in P} \zeta_t^p \right] \quad \forall t \in T
%     \end{aligned}\label{eq:2g}
% \end{equation}
% \begin{equation}
% \begin{aligned}
%     \sum_{ g \in G^b} (y^{g}_t + \eta^{g}_t) - (\delta^b_t + \zeta^b_t) +\psi^b_t
%     &= \sum_{ \forall u \in B^b} \left[ \Gamma^{bu} (\theta^b_t - \theta^u_t) \right], \forall t \in T
% \end{aligned} \label{eq:2e}
% \end{equation}

% \begin{equation}
%     \bar{p}_t = \sum_{b \in B} \delta_t^b  - \left[ \sum_{g \in G} (y^g_t +  \eta^{g}_t) \right]  \quad \forall t \in T.
%     \label{eq:20a}
% \end{equation}

In Equation \eqref{eq:2g}, the term \( \psi_t \) captures the total system demand deficit that should ideally be met. In Equation \eqref{eq:20a}, \( \bar{p}_t \) represents the net cracker-based power demand that remains to be fulfilled at the system level after accounting for contributions from conventional generation \( y^g_t \), renewable generation \( \eta^{g}_t \).
Next, we define the share of this target power demand attributable to each chemical plant microgrid \( p \in P \) through a normalized ratio as detailed in Equation \eqref{eq:20b}.
\begin{equation}
   \sum_{p \in P} r^p_t = 1, \text{ where, } r_t^p = \frac{\tilde{\zeta}_t^p}{\bar{p}_t} \quad \forall t \in T, \forall p \in P.
   \label{eq:20b}
\end{equation}
This ratio \( r_t^p \) expresses plant \( p \)'s requested demand $\tilde{\zeta}_t^p$ as a fraction of the residual system demand, providing a basis to assign a portion of the imbalance penalty back to the plant in a proportional way.
Using $r^p_t$, a modified penalty quantity \( q_t^p \) is constructed, which incorporates both the plant's own demand and its attributed share of system-level unmet demand as presented in Equation \eqref{eq:20c}.
\begin{equation}
    q_t^p = \tilde{\zeta}_t^p - (r_t^p \cdot \psi_t) \quad \forall t \in T,\, \forall p \in P.
    \label{eq:20c}
\end{equation}
This term \( q_t^p \) acts as a refined penalty signal for plant \( p \), considering not only its local demand but also its contribution to the broader system imbalance. Finally, we calculate the Lagrange multiplier \( \phi_t \) and update it at each iteration analogously to the dual update in \eqref{eq:19d}. This update penalizes the transmission system demand imbalance \( \psi_t \) defined in \eqref{eq:20c}.
\begin{equation}
    \phi_t \leftarrow \phi_t + \rho \cdot \psi_t
    \label{eq:20d}
\end{equation}
To incorporate these new formulations into the transmission system, we extend the augmented Lagrangian with the additional penalty term over the system-wide unmet demand as detailed in Equation \eqref{eq:21a}.
\begin{equation}\label{eq:21a}
\begin{aligned}
\min_{\zeta, f, \theta, x, y} & \mathcal{L}_{r}(\overline{\zeta}, \lambda, \phi) = \mathcal{L}_{r} + \sum_{t \in T} \left[ \phi_t\psi_t + \frac{\rho}{2} \| \psi_t \|^2 \right] + \sum_{b \in B} \sum_{t \in T} \bigg[ \lambda^{b}_{t} \left( \zeta_{t}^b - \overline{\zeta}_{t}^b \right) + \frac{\rho}{2} \left( \zeta_{t}^b - \overline{\zeta}_{t}^b \right)^2 \bigg]
\end{aligned}
\end{equation}

The unmet demand penalty serves to discourage persistent imbalances in \( \psi_t \), pushing the system toward more coordinated and feasible dispatch schedules. The total residual demand $\psi_{t,p}$, defined in \eqref{eq:2g}, is recomputed at each iteration using the updated plant demand values $\zeta_{t}^p$ and the current generation profile. In this expression, the left-hand side captures the total available supply from both conventional and renewable generators, while the right-hand side accounts for aggregate demand, including both bus-level loads and plant-specific electricity consumption.
% The total residual demand \( \psi_t \) is recalculated based on the updated plant demand values \( \zeta_t^p \) and the load/generation profile of the power grid:

% \begin{equation}
%     \psi_t = \left[ \sum_{g \in G} (y^g_t +  \eta^{g}_t) \right] - \left[ \sum_{b \in B} \delta_t^b + \sum_{p \in P} \zeta_t^p \right] \quad \forall t \in T.
%     \tag{21b}\label{eq:21b}
% \end{equation}

By incorporating this auxiliary system-level penalty into the model structure, the optimization formulation gains an additional mechanism to penalize global infeasibility and accelerate convergence. The added structure also helps guide each plant's local solver with more meaningful system-level information, all without violating data isolation since only consensus values are shared in the coordination process.

\subsection{Decentralized Chemical Plant Coordination}
To enable decentralized coordination between the power system and individual chemical plant microgrids, we employ ADMM at the plant level as well. ADMM facilitates the convergence of local plant-level decisions to a globally consistent solution by iteratively updating primal variables and dual variables. In this context, the coupling variable of interest is the cracker demand, which is denoted by \( \zeta_t^p \). For a particular plant $p$, the cracker power demand $\zeta_t^p$ is equal to $P^G_t$ as computed in equation \eqref{eq:gas_turbine} and is determined independently by each plant and reconciled at the system level through consensus.

At each iteration, the Lagrange multiplier \( \lambda^{p}_{t} \) is updated according to the difference between the plant-generated demand \( \tilde{\zeta}_t^p \) and the coordinated demand \( \dot{\zeta}_t^p \), as denoted in equation \eqref{eq:28}.
\begin{equation}
\lambda^{p}_{t} \gets \lambda^{p}_{t} + \rho \cdot \left( \tilde{\zeta}_t^p - \dot{\zeta}_t^p \right), \quad \forall t \in T.
\label{eq:28}
\end{equation}
This update penalizes deviations between the independently computed plant demand and the centralized consensus demand. The parameter \( \rho \) is a positive scalar that controls the strength of the penalty and, consequently, the rate of convergence. As the iterations progress, this dual update helps align the decentralized decisions with the centralized reference values.

To directly incorporate this coordination mechanism into the chemical plant microgrid optimization model, the Lagrangian penalty in \eqref{eq:29} is appended to the original microgrid objective in \eqref{eq:microgrid_obj}.
\begin{equation}\label{eq:29}
\begin{aligned}
\min_{\{P, F, x, M, \tilde\zeta\}} \; \mathcal{L}_{p}(\dot\zeta^p, \lambda^{p}) & = \mathcal{L}_p + \sum_{t \in T} \bigg[ \lambda^{p}_{t} (\dot{\zeta}_t^p - \tilde{\zeta}_t^p) + \frac{\rho}{2} (\dot{\zeta}_t^p - \tilde{\zeta}_t^p)^2 \bigg]
\end{aligned}
\end{equation}
The integration of the system-level penalty from the residual power demand into the chemical plant microgrid's local optimization follows a similar structure to the original ADMM-based coordination approach. However, with the introduction of residual penalty in Equations \eqref{eq:20c}, each chemical plant microgrid now receives an updated target demand \( q^{p}_{t} \) that reflects its share of the system-wide imbalance. This adjustment provides the plant with a more informed signal of where demand shortfalls exist and how its local decisions can help correct them.

As a result, the corresponding optimization problem solved by each microgrid at each iteration is extended to include both penalty terms, as depicted in equation \eqref{eq:31}.
\begin{equation}\label{eq:31}
\begin{aligned}
\min_{\{P, F, x, M, \tilde{\zeta}\}} \; \mathcal{L}_{p}(\dot{\zeta}^p, \lambda^{p}, \phi^{p}) &= \mathcal{L}_p + \sum_{t \in T} \left[ \lambda^{p}_{t} (\dot{\zeta}_t^p - \tilde{\zeta}_t^p) + \frac{\rho}{2} (\dot{\zeta}_t^p - \tilde{\zeta}_t^p)^2 \right] \\
+& \sum_{t \in T} \left[ \phi^{p}_{t} ( \tilde{\zeta}_t^p - q^{p}_{t}) + \frac{\rho}{2} (\tilde{\zeta}_{t}^p - q^{p}_{t})^2 \right]
\end{aligned}
\end{equation}

To incorporate this information into the decentralized optimization process, the Lagrange multiplier \( \phi^{p}_{t} \) of plant $p$ is now updated such that it accounts for the difference between local and consensus demand, alongside the relationship between the system-level coordination target \( q^{p}_{t} \) and the plant's proposed demand \( \tilde{\zeta}_t^p \), as given in equation \eqref{eq:30}.
\begin{equation}
\phi^{p}_{t} \gets \phi^{p}_{t} + \rho \cdot \left(\tilde{\zeta}_{t}^p -q^{p}_{t} \right), \quad \forall t \in T
\label{eq:30}
\end{equation}
This dual update mechanism encourages each plant to align with both the system-wide consensus and its proportional responsibility in meeting total grid demand. The term \( q^{p}_{t} \) effectively serves as a corrective guidance signal, nudging the plant toward load levels that improve global feasibility.

%Update local penalty $q_p^{k}$ via \eqref{eq:20c}\\
%Update local demand $\dot{\zeta}_p^{k}$ via \eqref{eq:19b} \;
\begin{algorithm}[!htbp]\label{alg:algorithm1}
\caption{Two-Phase Decentralized ADMM}
\SetAlgoNoLine
\DontPrintSemicolon
\SetInd{0.1em}{0.5em}
\SetKwComment{tcc}{$\triangleright$\ }{}
\KwIn{$\lambda_0,\, \zeta_0,\, \phi_0,\, x_0,\, y_0,\, \rho$;\enspace $\epsilon_0$}
\KwOut{$x^*,\, y^*,\, \theta^*,\, f^*$}
{
    $\pi \gets \mathrm{LP}$ \tcp*{PS LP Relaxation Phase}
    \For{$k = 1, 2, 3, \ldots$}{

        \tcc{CP Subproblem Solve}
        \For{$p \in \mathcal{P}$}{
            Solve Problem \eqref{eq:31} subject to \eqref{eq:power_balance}-\eqref{eq:ess_constraints}\\ %$\bf{z}^p_{k+1}\leftarrow argmin\; \mathcal{L}_\rho^p(\lambda_k,\, \phi_k)_{\pi}$\;
            Generate cracker demand estimate $\tilde{\zeta}^{p,k}$ via \eqref{eq:19b} \;
            Send demand estimate to PS: \texttt{MPI.Send}$(\{\zeta^{p,k}\})$
        }
        \BlankLine
        \tcc{PS Subproblem Sequence}
        Receive CP demand: \texttt{MPI.Recv}$(\{\tilde{\zeta}^{p,k}\})\ \forall p\in \mathcal{P}$\\
        \For{$b \in \mathcal{B}$}{
            Aggregate plant demands to bus level $\hat{\zeta}_{t}^{b,k}$ via \eqref{eq:19a}\;
            Compute consensus estimate for buses $\bar{\zeta}_{t}^{b,k}$ via \eqref{eq:19c}\;
        }
        Solve Problem \eqref{eq:21a} subject to \eqref{eq:2a}-\eqref{eq:2g}\\
        \For{$p \in \mathcal{P}$}{
            Generate bus-disaggregated cracker demand $\dot{\zeta}^{p,k}$ via \eqref{eq:19b} \;
            Generate system-level residual power penalty $q_p^{k}$ via \eqref{eq:20c}\\
        }
        %Calculate consensus variables $\{\tilde{\zeta}_p^{k},\, \tilde{q}_p^{k}\}$ via \eqref{eq:19c}\;
        Compute demand residual norm $\beta_{k} \leftarrow ||\boldsymbol{\zeta}_k - \tilde{\boldsymbol{\zeta}}_k||_2$\;
        Broadcast \texttt{MPI.Bcast}$(\{\dot{\zeta}^{p,k},\, q_p^{k}, \beta_k\}),\ \forall p\in \mathcal{P}$
        \BlankLine
        \tcc{Phase Transition and Convergence}
        \If{$\beta_k < \epsilon_0$ and $\pi = \mathrm{LP}$}{
            $\pi \leftarrow \mathrm{MIP}$ \tcp*{Integer Phase Change}
        }
        \If{$\beta_k < \epsilon_0$ and $\pi = \mathrm{MIP}$}{
                \textbf{break} \tcp*{Global Convergence}
        }

        \BlankLine
        \tcc{Dual Variable Updates at PS}
        Update $\lambda_{k+1,t}$ via \eqref{eq:19d} \ $\forall t \in T$ \;
        Update $\phi_{k+1,t}$ via \eqref{eq:20d} \ $\forall t \in T$ \;
        
        \BlankLine
        \tcc{Dual Variable Updates at CP}
        \For{$p \in \mathcal{P}$}{
            Update $\lambda^p_{k+1,t}$ via \eqref{eq:28} \ $\forall t \in T$\;
            Update $\phi^p_{k+1,t}$ via \eqref{eq:30} \ $\forall t \in T$ \;
        }
    }
}
\end{algorithm}
Together, these terms enhance the coordination between chemical plant microgrids and the power grid. By embedding both global and local perspectives into the plant's decision-making process, the model achieves faster convergence and greater robustness without violating the principle of decentralized privacy. The additional structure encourages each plant not only to match average expectations but also to contribute proportionally to system-wide demand satisfaction based on real-time imbalances.
\subsection{Joint Decentralized Operations Optimization Algorithm}
Algorithm \ref{alg:algorithm1} presents the decentralized coordination framework utilizing the Alternating Direction Method of Multipliers (ADMM) to reconcile chemical production constraints with power distribution dynamics. Our proposed ADMM-based decentralized coordination scheme leverages each computational agent $p \in \mathcal{P}$ to independently solve a cracker subproblem while iteratively exchanging consensus variables with the power system. The objective is to coordinate local optimization steps until global convergence is achieved. By enabling cracker plants to resolve local subproblems independently while exchanging boundary information, the system achieves a unified equilibrium. The methodology incorporates a specialized two-phase relaxation protocol with computation beginning with the Linear Programming (LP) mode which efficiently converges on a continuous trajectory. Since UC relaxations are tight \cite{padhy2004survey_uc}, the LP phase is applied only towards the power system formulation as a means to efficiently trace a continuous trajectory while effectively serving as a warm start for binary decisions. The algorithm automatically switches to a Mixed-Integer Programming (MIP) mode once a baseline tolerance is met to rigorously enforce discrete operational states.

The iterative coordination cycle proceeds through parallelizing chemical plant microgrid and transmission system ADMM updates. At each iteration, chemical plant microgrids minimize their respective augmented Lagrangians to determine energy needs, subject to internal physical limits, and transmit their estimates to the transmission system stakeholder. The transmission system then aggregates these inputs to resolve the transmission dispatch problem, calculating system-wide consensus targets. The power system solves its own local problem with the given consensus estimates and computes the bus-disaggregated consensus estimates as well as the target residual demand penalty to each chemical plant.
These optimized targets and consensus estimates are broadcast back to the individual plants, accompanied by an error metric $\beta_k$ that tracks alignment with convergence thresholds. The process is sustained by a dual update mechanism that systematically adjusts Lagrangian multipliers based on the mismatch between decentralized decisions and global requirements. The algorithm persists through this cycle until the MIP-mode residuals fall below a specified threshold, yielding the optimal dispatch configuration and guaranteeing global optimality.

Using the decentralized joint operations optimization formulation, we now present our experimental results for large-scale integrated cracker demand and power systems operations.

\section{Experimental Results} \label{sec:results}

To characterize the cost of privacy-friendly decomposition on joint power and chemical system decisions, we design a comparative experimental setup in which both a centralized integrated formulation and the proposed decentralized framework are applied to a curated test case comprising 26 ethylene plants from Texas coupled with the ACTIVSg2000 synthetic Texas transmission network \cite{activsg2000_testcase,matpower}. The locations, feedstock compositions, and ethylene production capacities of the ethylene plants were obtained using data meticulously curated from several sources, including news organizations \cite{sabicnews}, open and closed-source market analysis reports \cite{Precedence2024}, and academic literature \cite{crackingsurvey}. The centralized model, which assumes full information sharing between the grid operator and all chemical plant microgrids, serves as the reference point against which the consequences of data isolation are measured. Our experiments were conducted using a 24-hour operational planning horizon. Chemical plant microgrids were mapped to their nearest buses based on geographic proximity.

\subsection{Centralized Benchmark Design}

In the centralized experiment, the chemical plant microgrid demand is treated as an exogenous parameter within the power system optimization model. The coordination between the grid and the plants is inherently embedded in a single integrated optimization problem, meaning that the system operator has full visibility into both power system operations and plant electricity demand. Our centralized model is solved using Gurobi 10.0.3, and serves as the baseline for both objective value and computational time. The centralized framework assumes no information barriers and reflects an ideal scenario in terms of global optimality and coordination. However, it does not capture the reality of operational privacy or modular decision-making present in large-scale decentralized systems. The results from this setup are used to quantify the impact of decentralization on optimality and runtime.

\subsection{Decentralized Experimental Implementation}

The decentralized setup implements our proposed ADMM based coordination mechanism in this paper, in which the power system and each chemical plant microgrids are modeled as independent agents that solve their respective subproblems and coordinate via an ADMM-based iterative scheme. Communication between agents is handled using the Message Passing Interface (MPI) \cite{dalcin2008mpi4py}, with the Python interface \verb|mpi4py| used to facilitate inter-process communication. This design allows for a scalable, modular execution of the decentralized algorithm across a high-performance computing (HPC) cluster. Each of the 26 microgrids and the centralized power grid operator are treated as separate processes, resulting in 27 total compute nodes. These experiments were conducted on an HPC environment consisting of Intel Skylake compute nodes, ensuring sufficient memory and parallel processing capability to simulate real-world large-scale systems.

In each ADMM iteration, plants solve their local optimization problems and communicate cracker demand information to the grid process. The power grid process aggregates this information, solves its own subproblem, and updates consensus values and dual variables, which are then sent back to the plants for the next iteration. ADMM iterations continue until convergence criteria based on primal and dual residuals are satisfied. The decentralized experiments use the same data, parameters, and initial conditions as the centralized case to ensure comparability. Results are analyzed in terms of solution optimality, computational time, and convergence behavior enabling a clear understanding of the trade-offs between global coordination and modular, privacy-friendly optimization.

\subsection{Comparative Study of Centralized and Decentralized Planning Methodologies}

We evaluate the performance of the decentralized optimization framework by comparing it against the centralized benchmark across a range of power system demand levels. Three demand scenarios, Low, Medium, and High, were considered, each evaluated under two levels of electrification from the ethylene plants (10\% and 50\%). The centralized model represents an idealized case with complete visibility and control over all system components, while the decentralized model reflects a more realistic architecture where ethylene plants and the power grid operate as separate agents, coordinated through the proposed ADMM-based method. Tables~\ref{tab:cost_comparison_10} and~\ref{tab:cost_comparison_50} summarize the comparison between centralized and decentralized results in terms of total cost at 10\% and 50\% electrification, respectively, broken down into Unit Commitment cost (UC), Dispatch cost (DC), Demand Curtailment cost (DCC), and Chemical Plant cost (CPC). Each table also reports the decentralized accuracy metrics: 2-norm consensus error and optimality gap.

% Across all test cases, the largest observed optimality gap is under 1.8\%, directly quantifying the cost of data isolation on joint power and chemical system decisions. Notably, demand curtailment is observed in the decentralized cases, as a direct consequence of plants operating without full visibility into the grid's state, forcing demand coordination through consensus signals alone rather than direct observation. These results characterize the privacy-optimality tradeoff inherent to decomposition-based coordination in coupled power-chemical energy systems.

\begin{table*}[htbp]
\caption{Normalized costs (\$ $\times$ 10$^6$) and decentralized accuracy metrics for 10\% electrification.}
\centering
\normalsize
\resizebox{\textwidth}{!}{%
\begin{tabular}{lcccccccccccc}
\toprule
\multirow{2}{*}{\textbf{Demand profile}}
& \multicolumn{2}{c}{\textbf{UC}}
& \multicolumn{2}{c}{\textbf{DC}}
& \multicolumn{2}{c}{\textbf{DCC}}
& \multicolumn{2}{c}{\textbf{CPC}}
& \multicolumn{2}{c}{\textbf{Total}}
& \multirow{2}{*}{\textbf{2-norm error}}
& \multirow{2}{*}{\textbf{Opt. gap}} \\
\cmidrule(lr){2-3}
\cmidrule(lr){4-5}
\cmidrule(lr){6-7}
\cmidrule(lr){8-9}
\cmidrule(lr){10-11}
& \textbf{Cent.} & \textbf{Decent.}
& \textbf{Cent.} & \textbf{Decent.}
& \textbf{Cent.} & \textbf{Decent.}
& \textbf{Cent.} & \textbf{Decent.}
& \textbf{Cent.} & \textbf{Decent.}
& & \\
\midrule
\textbf{Low}    & 0.144 & 1.126 & 7.547  & 6.553  & 0 & 0.043 & 3.882 & 3.920 & 11.575 & 11.599 & 7.54 & 0.224\% \\
\textbf{Medium} & 1.534 & 2.547 & 14.559 & 13.229 & 0 & 0.015 & 3.907 & 4.309 & 20.000 & 20.099 & 4.94 & 0.493\% \\
\textbf{High}   & 2.962 & 3.918 & 19.408 & 18.191 & 0 & 0.013 & 3.967 & 4.253 & 26.336 & 26.375 & 4.83 & 0.144\% \\
\bottomrule
\end{tabular}%
}
\label{tab:cost_comparison_10}
\end{table*}

\vspace{-1em}

\begin{table*}[htbp]
\caption{Normalized costs (\$ $\times$ 10$^6$) and decentralized accuracy metrics for 50\% electrification.}
\centering
\normalsize
\resizebox{\textwidth}{!}{%
\begin{tabular}{lcccccccccccc}
\toprule
\multirow{2}{*}{\textbf{Demand profile}}
& \multicolumn{2}{c}{\textbf{UC}}
& \multicolumn{2}{c}{\textbf{DC}}
& \multicolumn{2}{c}{\textbf{DCC}}
& \multicolumn{2}{c}{\textbf{CPC}}
& \multicolumn{2}{c}{\textbf{Total}}
& \multirow{2}{*}{\textbf{2-norm error}}
& \multirow{2}{*}{\textbf{Opt. gap}} \\
\cmidrule(lr){2-3}
\cmidrule(lr){4-5}
\cmidrule(lr){6-7}
\cmidrule(lr){8-9}
\cmidrule(lr){10-11}
& \textbf{Cent.} & \textbf{Decent.}
& \textbf{Cent.} & \textbf{Decent.}
& \textbf{Cent.} & \textbf{Decent.}
& \textbf{Cent.} & \textbf{Decent.}
& \textbf{Cent.} & \textbf{Decent.}
& & \\
\midrule
\textbf{Low}    & 0.202 & 1.167 & 7.902  & 7.097  & 0 & 0.004 & 6.703 & 6.813 & 14.808 & 14.837 & 7.89 & 1.791\% \\
\textbf{Medium} & 1.611 & 2.677 & 14.917 & 13.846 & 0 & 0.006 & 6.727 & 6.806 & 23.245 & 23.335 & 3.20 & 0.386\% \\
\textbf{High}   & 3.067 & 4.068 & 19.761 & 18.705 & 0 & 0.008 & 6.779 & 6.806 & 29.607 & 29.618 & 3.19 & 0.037\% \\
\bottomrule
\end{tabular}%
}
\label{tab:cost_comparison_50}
\end{table*}

Across all test cases, the largest observed optimality gap is under 1.8\%, directly quantifying the cost of data isolation on joint power and chemical system decisions. Notably, demand curtailment is observed in the decentralized cases, as a direct consequence of plants operating without full visibility into the grid's state, forcing demand coordination through consensus signals alone rather than direct observation. These results characterize the privacy-optimality tradeoff inherent to decomposition-based coordination in coupled power-chemical energy systems.

\subsection{Analysing Convergence Trends}
An important aspect of evaluating the performance of the decentralized optimization framework is its convergence behavior across different coordination strategies. In this section, we examine how each method converges over successive iterations by tracking the change in demand coordination throughout the ADMM process using 2-norm residual errors.
\begin{figure*}[!t]
    \centering

    % Left column: 10% electrification
    \begin{minipage}{0.48\textwidth}
        \centering
        \subfigure[Low demand, 10\% electrification]{
            \includegraphics[width=\linewidth]{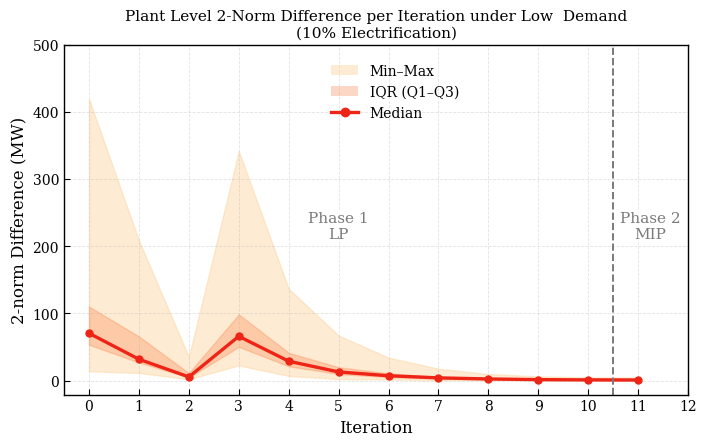}
            \label{fig:low_10pct_error}
        }

        \vspace{0.25cm}

        \subfigure[Medium demand, 10\% electrification]{
            \includegraphics[width=\linewidth]{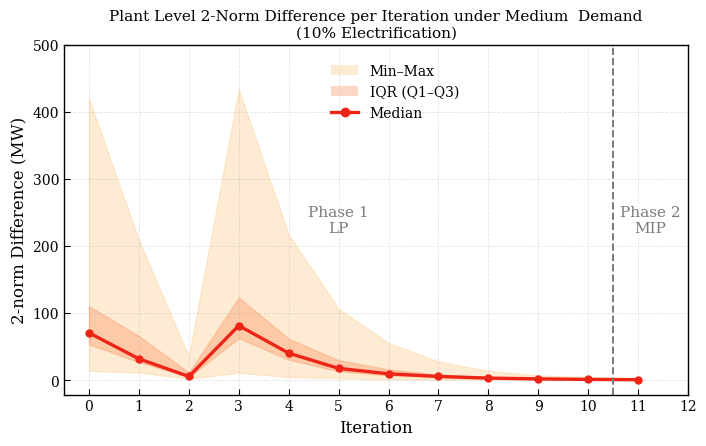}
            \label{fig:med_10pct_error}
        }

        \vspace{0.25cm}

        \subfigure[High demand, 10\% electrification]{
            \includegraphics[width=\linewidth]{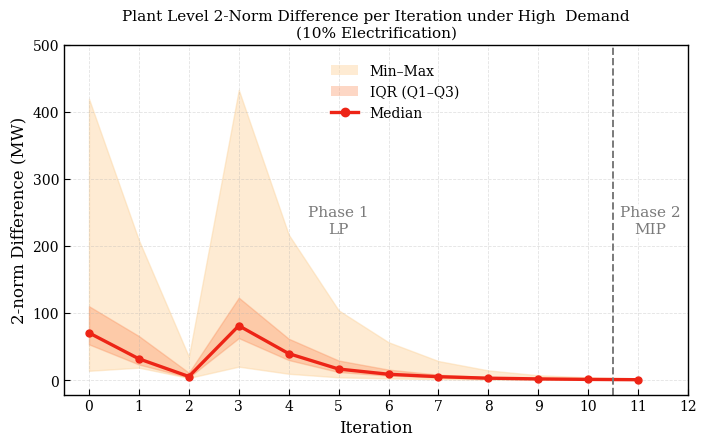}
            \label{fig:high_10pct_error}
        }
    \end{minipage}
    \hfill
    % Right column: 50% electrification
    \begin{minipage}{0.48\textwidth}
        \centering
        \subfigure[Low demand, 50\% electrification]{
            \includegraphics[width=\linewidth]{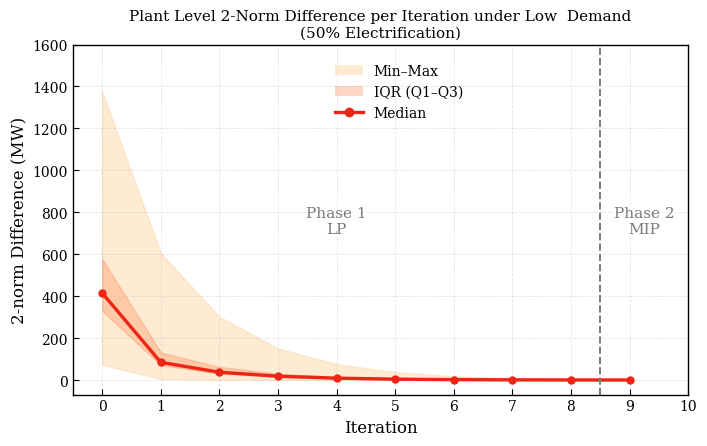}
            \label{fig:low_50pct_error}
        }

        \vspace{0.25cm}

        \subfigure[Medium demand, 50\% electrification]{
            \includegraphics[width=\linewidth]{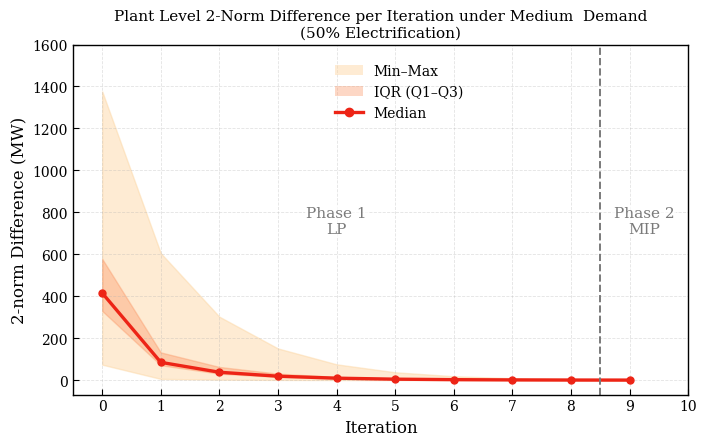}
            \label{fig:med_50pct_error}
        }

        \vspace{0.25cm}

        \subfigure[High demand, 50\% electrification]{
            \includegraphics[width=\linewidth]{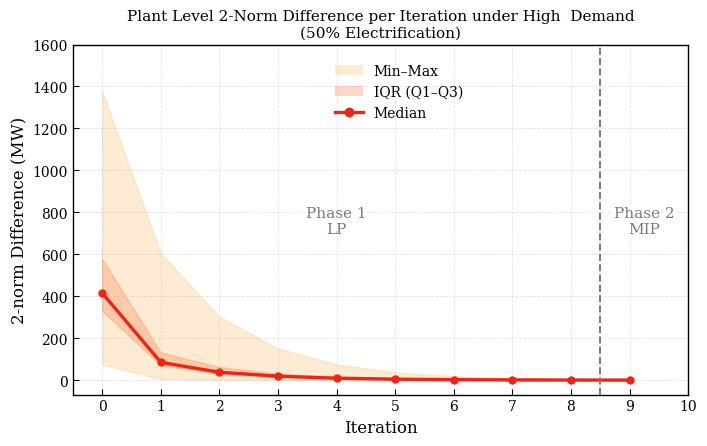}
            \label{fig:high_50pct_error}
        }
    \end{minipage}

    \caption{ADMM convergence, measured by 2-norm difference versus iteration, under low, medium, and high power system demand scenarios. The left column represents 10\% electrification, while the right column represents 50\% electrification.}
    \label{fig:convergence}
\end{figure*}

A notable distinction emerges between the convergence profiles at 10\% and 50\% electrification. At 50\% electrification presented in Figures ~\ref{fig:low_50pct_error}--~\ref{fig:high_50pct_error}, the 2-norm error exhibits a smooth, monotonic decline across all demand scenarios, reaching consensus within approximately 8 LP iterations before the final MIP solve. In contrast, the 10\% electrification cases represented in Figures~\ref{fig:low_10pct_error}--~\ref{fig:high_10pct_error} display a characteristic oscillation where the 2-norm error decreases sharply during the first two iterations, then spikes upward around iteration 3 before resuming its downward trajectory. This non-monotone behavior is a direct consequence of the asymmetric binary treatment in the two-phase ADMM heuristic. During Phase 1, only the power system commitment variables ($x_{g,t}$) are relaxed to continuous $[0,1]$, while the chemical plant microgrid binaries (dispatchable unit commitments ($x^D_{i,t}$), fuel cell status ($x^{FC}_t$), and battery charge/discharge states ($x^{ESS}_{C,t}$, $x^{ESS}_{DC,t}$)) remain strictly integer throughout. As the ADMM penalty signals update between iterations, these plant-side binaries can flip discretely such as a generator switching from off to on, or a battery transitioning from charging to discharging. 

Consequently, these binary flips can cause discontinuous jumps in the plant's grid power demand ($P^G_t$) and resulting in sudden increases in the 2-norm error. At 10\% electrification, the electric cracker demand is relatively small, leaving the plant's internal dispatch with greater scheduling flexibility. As a result, the local solver has more room to toggle binaries between feasible configurations across iterations. At 50\% electrification, the substantially larger fixed electric cracker demand ($P_{EC}$) constrains the plant's internal degrees of freedom, effectively anchoring most internal generators and storage systems into committed states, which suppresses binary oscillation and yields the smoother convergence profile observed. Despite the transient oscillation at 10\%, the framework ultimately converges to tight consensus in all cases, with the final 2-norm errors and optimality gaps remaining comparable across both electrification levels.

\subsection{Computational Performance of the Two Phase Approach}
\begin{figure*}[!t]
    \centering
    \subfigure[Timing Breakdown for 10\% Electrification]{
        \includegraphics[width=0.47\textwidth]{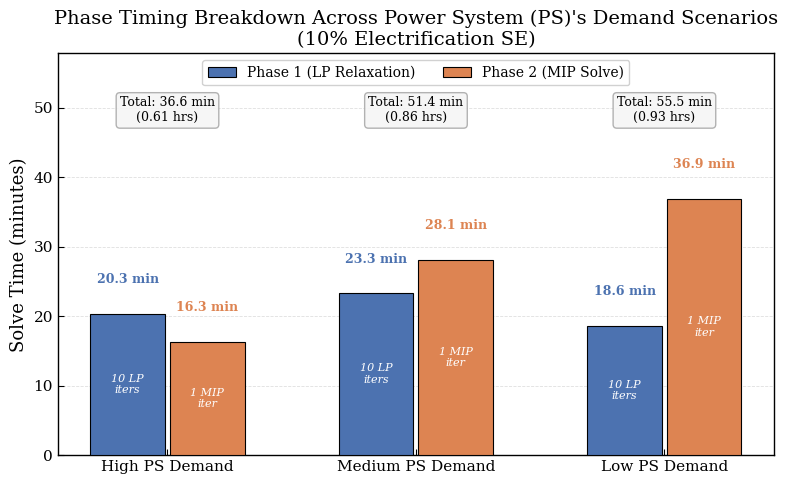}
        \label{fig:timing_10_pct}
    }
    \hfill
    \subfigure[Timing Breakdown for 50\% Electrification]{
        \includegraphics[width=0.47\textwidth]{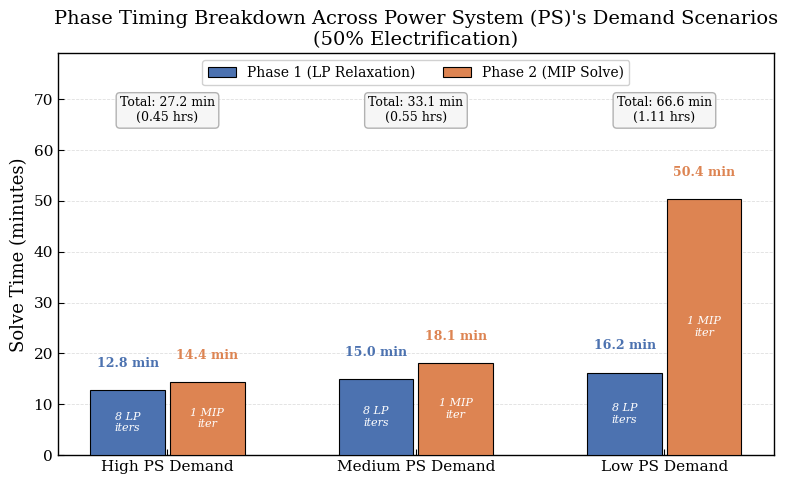}
        \label{fig:timing_50_pct}
    }
    \caption{Phase timing breakdown across power system demand scenarios for 10\% and 50\% electrification levels, showing Phase~1 (LP relaxation) and Phase~2 (MIP solve) contributions.}
    \label{fig:timing_breakdown}
\end{figure*}

Figure~\ref{fig:timing_breakdown} presents the wall-clock timing breakdown across all six experimental configurations, decomposed into Phase~1 (LP relaxation) and Phase~2 (MIP solve) contributions. Across both electrification levels, Phase~1 exhibits relatively stable computation times, ranging from 12.8 to 23.3~minutes regardless of demand scenario. This consistency arises because the LP-relaxed subproblems have similar computational complexity once the integer constraints are removed; the primary variation stems from the number of ADMM iterations required to reach the convergence threshold of $\epsilon = 10$. At 10\% electrification, all three demand scenarios require 11 iterations to converge, while at 50\% electrification, convergence is achieved in 9 iterations. The faster convergence at higher electrification is attributable to the larger coupling demands providing a stronger signal for the ADMM consensus mechanism, enabling the power system and chemical plant microgrid subproblems to reach agreement more rapidly.

In contrast, Phase~2 exhibits substantially greater variability. The single MIP iteration, which restores all binary commitment variables and solves the full mixed-integer program, accounts for the majority of solve time in four of the six cases. At 10\% electrification, Phase~2 ranges from 16.3~minutes (High demand) to 36.9~minutes (Low demand). At 50\% electrification, this disparity is more pronounced with the High demand scenario completing in 14.4~minutes, while the Low demand scenario requires 50.4~minutes, which happens to be 3.5 times longer. This behavior reflects the combinatorial difficulty of the UC problem under low-demand conditions, where fewer generators are needed but the selection among candidate units creates a harder branch-and-bound landscape for the solver.

Total wall-clock times range from 27.2~minutes (50\%, High demand) to
66.6~minutes (50\%, Low demand), with most configurations completing in under one hour. Notably, all chemical plant subproblems solve in under 2~seconds per iteration across every configuration, confirming that the computational bottleneck resides entirely in the power system MIP solve rather than in the plant-level optimization. These results demonstrate that the two-phase LP-MIP strategy effectively limits the number of expensive integer solves to a single final iteration, keeping total runtimes practical for day-ahead operational planning.

\subsection{Decarbonized Grid Results}
As the transition toward low-carbon energy systems accelerates, it is critical to evaluate not only economic performance but also environmental impact. This section presents a carbon accounting framework that quantifies \ce{CO2} emissions across the centralized and decentralized configurations, using power-system dispatch results, generator fuel types, and emission intensity factors. We report emissions at the \emph{transmission system level}, which captures the total carbon footprint of serving all grid demand, including both normal consumer bus load and the electrified chemical-plant load. The primary objective of this analysis is to assess how varying levels of chemical-plant electrification influence carbon emissions on the transmission system side under a decentralized solution paradigm. As a result, we compute the total emissions estimates based on the share of generation fuel on the transmission side and derive a generation mix ratio for calculating effective electrified cracker-based emissions as a consequence of grid integration.

%Further we seek to also establish whether decentralized coordination can support decarbonization goals without sacrificing operational performance.

\noindent \textit{Emission Contribution Owing to Electrified Cracker Integration}: Emissions on the transmission side arise only from the coal and natural-gas generators; nuclear and renewable generation are carbon-free \cite{activsg2000_testcase}. Using the fuel-specific emission factors $\alpha_\text{coal}$ and $\alpha_\text{ng}$ derived from the Energy Information Administration \cite{eia}, the emissions from each fossil source can be represented by Equation \eqref{eq:27a} where $P_g$ is the dispatched power of generator $g$.
\begin{equation}\label{eq:27a}
    \text{\ce{CO2}}_\text{coal} = \sum_{\forall g \in G_\text{coal}} \alpha_\text{coal} \cdot P_g,\ \text{\ce{CO2}}_\text{ng} = \sum_{\forall g \in G_\text{ng}} \alpha_\text{ng} \cdot P_g 
\end{equation}
The system-wide emission is consequently denoted by equation \eqref{eq:co2_sys} which captures the total carbon footprint of the grid under a given dispatch.
\begin{equation}
    \text{\ce{CO2}}_\text{sys}
    = \text{\ce{CO2}}_\text{coal} + \text{\ce{CO2}}_\text{ng}.
    \label{eq:co2_sys}
\end{equation}
\noindent The carbon accounting in \eqref{eq:27a}--\eqref{eq:co2_sys} lets us compare the centralized and decentralized architectures on environmental grounds with varying transmission demand and cracker electrification levels. Because the chemical-plant load is served from the same grid, changes in plant electrification shift the dispatch mix and are reflected directly in $\text{\ce{CO2}}_\text{sys}$.
% \noindent \textit{Generation Mix Ratios}: We know that the total power generated on the transmission side is the sum of coal, natural-gas, and renewable contributions as detailed in equation \eqref{eq:total_power}.
% \begin{equation}
%     P_\text{grid}^\text{total} = P_\text{grid}^\text{coal} + P_\text{grid}^\text{NG} + P_\text{grid}^\text{RW}.
%     \label{eq:total_power}
% \end{equation}
% The share of coal and natural gas as a fraction of total generation on the transmission side are given by equation \eqref{eq:ratios} where $\gamma_\text{grid}^\text{FF}$ is the fossil-fuel share of generation, a compact indicator of dispatch emissions. 
% \begin{gather}
%     \gamma_\text{grid}^\text{coal} = \frac{P_\text{grid}^\text{coal}}{P_\text{grid}^\text{total}},\quad
%     \gamma_\text{grid}^\text{NG}   = \frac{P_\text{grid}^\text{NG}}{P_\text{grid}^\text{total}},\quad
%     \gamma_\text{grid}^\text{FF}   = \frac{P_\text{grid}^\text{coal}+P_\text{grid}^\text{NG}}{P_\text{grid}^\text{total}},
%     \label{eq:ratios}
% \end{gather}

% \noindent \textit{Cracker demand emissions at plants arising from grid supplied energy}: 

% \begin{gather}
%     \alpha_\text{coal}\cdot\gamma_\text{grid}^\text{coal}\sum\limits_{p\in\mathcal{P}}\zeta^{*p}_t + \alpha_\text{ng}\cdot\gamma_\text{grid}^\text{NG}\sum\limits_{p\in\mathcal{P}}\zeta^{*p}_t 
% \end{gather}

% ---- System-level emissions results table ----
\begin{table*}[htbp]
\caption{Comparison of grid emissions arising from various levels of cracker electrification (in tons).}
\centering
\normalsize
\resizebox{\textwidth}{!}{%
\begin{tabular}{lrrrrrr}
\toprule
\multirow{2}{*}{\textbf{Electrification}}
& \multicolumn{3}{c}{\textbf{High demand}}
& \multicolumn{3}{c}{\textbf{Low demand}} \\
\cmidrule(lr){2-4}
\cmidrule(lr){5-7}
& \textbf{Centralized}
& \textbf{Decentralized}
& \textbf{Difference}
& \textbf{Centralized}
& \textbf{Decentralized}
& \textbf{Difference} \\
\midrule
10\% & 607{,}340 & 612{,}137 & 0.784\%
     & 234{,}871 & 236{,}166 & 0.548\% \\
50\% & 626{,}141 & 626{,}339 & 0.032\%
     & 256{,}743 & 255{,}575 & $-0.457$\% \\
\bottomrule
\end{tabular}%
}
\label{tab:emissions_sidebyside}
\end{table*}
Table~\ref{tab:emissions_sidebyside} reports system-level \ce{CO2} for both architectures. Across every case, the decentralized framework reproduces the centralized emissions to within approximately 0.8\%, and to within 0.03\% at 50\% electrification under high demand. This close agreement demonstrates that decentralized coordination, despite sharing only limited information between the power system and the ethylene plant microgrids, attains essentially the same generation mix and therefore the same carbon footprint as the fully observable centralized benchmark. Emissions scale primarily with power-system demand, and increase modestly with electrification as additional grid power is drawn to serve the larger electrified cracker load. Crucially, the persistence of near-identical centralized and decentralized emissions across all demand and electrification levels indicates that privacy-preserving decentralization does not come at an environmental cost: the data-isolation mechanism that makes industrial integration acceptable to plant stakeholders carries no measurable carbon penalty relative to a fully centralized solution. This result directly advances the decarbonization objective of this work, by coordinating their microgrids with the grid through minimal, privacy-respecting signals, electrified chemical plants can draw on an increasingly renewable power supply and inherit its emissions reductions directly, without disclosing proprietary operational data. As the grid continues to decarbonize, our framework offers a practical pathway for industrial decarbonization that remains compatible with real-world data-governance constraints.

% Table~\ref{tab:emissions_sidebyside} reports system-level \ce{CO2} for both
% architectures. Across every case, the decentralized framework reproduces the
% centralized emissions to within approximately 0.8\%, and to within 0.03\% at
% 50\% electrification under high demand. This close agreement demonstrates
% that decentralized coordination, despite sharing only limited information
% between the power system and the chemical plants, attains essentially the
% same generation mix, and therefore the same carbon footprint, as the fully
% observable centralized benchmark. Emissions rise modestly with both demand
% and electrification, consistent with the additional fossil generation needed
% to serve the larger electrified load. The persistence of near-identical
% centralized and decentralized emissions across all demand and electrification
% levels indicates that privacy-preserving decentralization does not come at an
% environmental cost: the framework supports decarbonization objectives without
% sacrificing the coordination quality of a centralized solution.

\section{Conclusion} \label{sec:conclusion}
In this paper, we characterize the cost of privacy-friendly decomposition on joint Unit Commitment and chemical plant microgrid decisions, contributing one of the first systematic analyses of the privacy-optimality tradeoff in coupled power transmission and chemical plant systems in the context of electrification for decarbonization. Unlike a fully centralized approach that requires extensive information exchange, our method isolates the data of each system by sharing only minimal coordination signals. We achieve this through the Alternating Direction Method of Multipliers (ADMM), where each subsystem solves its own local subproblem and iteratively updates a global consensus estimate via penalty functions and Lagrangian multipliers. To further strengthen convergence and reduce computational overhead, we introduce auxiliary system-level penalties that refine local coordination.

Through numerical experiments on a geographically grounded Texas grid case study comprising 26 ethylene plants coupled with the ACTIVSg2000 synthetic transmission network, we demonstrate that data isolation incurs a modest optimality penalty, with gaps remaining below 1.8\% across all tested configurations, while introducing demand curtailment as a direct consequence of limited grid visibility. Computationally, the proposed two-phase LP-to-MIP strategy limits expensive integer solves to a single final iteration demonstrating robustness of the two phase approach. We also identify a mechanistic link between electrification level and convergence behavior where plant-side binary commitment variables exhibit inter-iteration flipping that produces oscillatory convergence profiles at lower electrification (10\%). On the other hand, higher electrification (50\%), larger fixed electric-cracker loads anchor these binaries and yield smooth, monotonic convergence. On the environmental front, the decentralized framework reproduces centralized system-level CO$_2$ emissions to within approximately 0.8\% across all scenarios, demonstrating that privacy-preserving decentralization carries no measurable carbon penalty relative to a fully observable benchmark.

These findings underscore the real-world viability of privacy-friendly decomposition for industrial energy systems that demand both operational confidentiality and high-quality optimization outcomes. While our results rely on a synthetic transmission network and a deterministic 24-hour planning horizon, the framework offers a replicable foundation for future work incorporating stochastic renewable uncertainty, dynamic locational marginal pricing signals, and validation on real-world grid data. As chemical process electrification accelerates, such decomposed coordination mechanisms will be essential for enabling industrial decarbonization pathways that remain compatible with real-world data-governance constraints.

% Elsevier-required declarations 

%\section*{CRediT authorship contribution statement}

\section*{CRediT authorship contribution statement}
\noindent Richard Reed: Data Curation, Conceptualization, Methodology, Software, Formal analysis, Investigation, Validation, Writing - original draft, Writing - review \& editing. Kazi Arman Ahmed: Conceptualization, Methodology, Software, Formal analysis, Investigation, Validation, Writing - original draft, Writing - review \& editing, Visualization. Saba Ghasemi: Data curation, Conceptualization, Methodology, Software, Resources, Writing - review \& editing. Zheyu Jiang: Data curation, Validation, Writing - review \& editing, Supervision, Project administration, Funding acquisition. Paritosh Ramanan: Conceptualization, Methodology, Software, Formal analysis, Writing - review \& editing, Supervision, Project administration, Funding acquisition.

\section*{Funding}
\noindent This work was supported by the National Science Foundation (NSF) award number 2343072. The authors acknowledge the support under the EAGER: CET: Decentralized Algorithms for Integrating Decarbonized Chemical Process Heating with Renewable-driven, Electric Power Systems program.

\section*{Declaration of competing interest}
None

\section*{Data availability}
Data will be made available on request.

\bibliographystyle{elsarticle-num}
\bibliography{main}

\end{document}